\documentclass[twocolumn,showpacs,preprintnumbers,amsmath,amssymb,floatfix]{revtex4}

\usepackage{graphicx}
\usepackage{dcolumn}
\usepackage{bm}

\begin{document}

%

\let\a=\alpha      \let\b=\beta       \let\c=\chi        \let\d=\delta
\let\e=\varepsilon \let\f=\varphi     \let\g=\gamma      \let\h=\eta
\let\k=\kappa      \let\l=\lambda     \let\m=\mu
\let\o=\omega      \let\r=\varrho     \let\s=\sigma
\let\t=\tau        \let\th=\vartheta  \let\y=\upsilon    \let\x=\xi
\let\z=\zeta       \let\io=\iota      \let\vp=\varpi     \let\ro=\rho
\let\ph=\phi       \let\ep=\epsilon   \let\te=\theta
\let\n=\nu
\let\D=\Delta   \let\F=\Phi    \let\G=\Gamma  \let\L=\Lambda
\let\O=\Omega   \let\P=\Pi     \let\Ps=\Psi   \let\Si=\Sigma
\let\Th=\Theta  \let\X=\Xi     \let\Y=\Upsilon

%

%

\def\cA{{\cal A}}                \def\cB{{\cal B}}
\def\cC{{\cal C}}                \def\cD{{\cal D}}
\def\cE{{\cal E}}                \def\cF{{\cal F}}
\def\cG{{\cal G}}                \def\cH{{\cal H}}
\def\cI{{\cal I}}                \def\cJ{{\cal J}}
\def\cK{{\cal K}}                \def\cL{{\cal L}}
\def\cM{{\cal M}}                \def\cN{{\cal N}}
\def\cO{{\cal O}}                \def\cP{{\cal P}}
\def\cQ{{\cal Q}}                \def\cR{{\cal R}}
\def\cS{{\cal S}}                \def\cT{{\cal T}}
\def\cU{{\cal U}}                \def\cV{{\cal V}}
\def\cW{{\cal W}}                \def\cX{{\cal X}}
\def\cY{{\cal Y}}                \def\cZ{{\cal Z}}
%

\newcommand{\Ns}{N\hspace{-4.7mm}\not\hspace{2.7mm}}
\newcommand{\qs}{q\hspace{-3.7mm}\not\hspace{3.4mm}}
\newcommand{\ps}{p\hspace{-3.3mm}\not\hspace{1.2mm}}
\newcommand{\ks}{k\hspace{-3.3mm}\not\hspace{1.2mm}}
\newcommand{\des}{\partial\hspace{-4.mm}\not\hspace{2.5mm}}
\newcommand{\desco}{D\hspace{-4mm}\not\hspace{2mm}}



\title{ Four Generation $CP$ Violation in
 $B \to \phi K^0$, $\pi^0 K^0$, $\eta^\prime K^0$ and Hadronic Uncertainties
}
\author{Wei-Shu Hou$^{a}$}
\email{wshou@phys.ntu.edu.tw}
\author{Makiko Nagashima$^a$}
\email{makiko@phys.ntu.edu.tw}
\author{Guy Raz$^{b}$}
\email{guy.raz@weizmann.ac.il}
\author{Andrea Soddu$^{b}$}
\email{andrea.soddu@weizmann.ac.il}
\affiliation{ $^a$Department of Physics, National Taiwan
 University, Taipei, Taiwan 10617, R.O.C. \\
$^b$Department of Particle Physics, Weizmann Institute
 of Science, Rehovot 76100, Israel
}
\date{\today}

\begin{abstract}
The fourth generation can give the correct trend of ${\cal
S}_{\phi K^0}$, ${\cal S}_{\pi^0 K^0} < \sin2\phi_1/\beta$, as
indicated by data, and the effect, being largely leading order, is
robust against hadronic uncertainties. The effect on ${\cal
S}_{\eta' K^0}$, however, is diluted away by hadronic effects, and
${\cal S}_{\eta' K^0} \simeq \sin2\phi_1/\beta$ is expected. The
near maximal $\arg V^*_{t's}V_{t'b} \lesssim 90^\circ$ that is
needed could resolve the unequal direct $CP$ violation seen in
$B\to K^+\pi^-$ and $K^+\pi^0$ modes, and is consistent with $b\to
s\ell^+\ell^-$ and $B_s$ mixing constraints.
\end{abstract}

\maketitle


At present there are two hints for possible New Physics (NP) from
$CP$ violation (CPV) studies in the $B$ system, both in charmless
$b\to s$ transitions.

Time dependent CPV (TCPV) in $B$ decays to $CP$ eigenstate $f$ is
measured by
\begin{eqnarray}
&&
 \frac{\Gamma(\overline{B}^0_{}(t)\to f) - \Gamma(B^0_{}(t)\to f)}
 {\Gamma(\overline{B}^0_{}(t)\to  f) + \Gamma(B^0_{}(t)\to f)}
 \nonumber \\
 &=& S_f\sin(\Delta m_B\, t) -C_f\cos(\Delta m_B\, t).
\end{eqnarray}
We concern ourselves with $S_f$ only, since $-C_f = A_f$ are all
consistent with zero so far.
The current world average of $S_f$ in $b\to c\bar cs$ decays gives
$\sin2\phi_1/\beta = 0.69 \pm 0.03$~\cite{HFAG}, which dominantly
comes from $B^0\to J/\psi K_S$.
TCPV measurements in loop dominated $b\to s\bar qq$ processes such
as $B^0 \to \eta^\prime K^0$, $\phi K^0$ and $\pi^0 K^0$, on the
other hand, have persistently given values below
$\sin2\phi_1/\beta$. The current values~\cite{HFAG} are ${\cal
S}_{\eta^\prime K^0} = 0.50 \pm 0.09$ \cite{etapK}, ${\cal
S}_{\phi K^0} = 0.47 \pm 0.19$ and ${\cal S}_{\pi^0 K^0} = 0.31
\pm 0.26$.

It was suggested some time ago~\cite{btosqqNP} that, for $b\to s\bar
qq$ final states, a significant $\Delta {\cal S}_f = {\cal S}_f -
\sin2\phi_1/\beta$ would indicate NP. The study of theoretical
uncertainties for $\Delta {\cal S}_f$ has therefore been a great focus
during the past year. Consensus has emerged that $\Delta {\cal S}_f$
in these modes tend to be small and {\it
  positive}~\cite{Beneke,CCS,BHNR} within the Standard Model (SM),
which is opposite the trend seen by experiment. It is therefore
imperative to establish $\Delta {\cal S}_f \neq 0$ experimentally
beyond any doubt in a few modes, which would require considerably more
data than present.

A simpler measurement than ${\cal S}_f$ is direct CPV (DCPV)
asymmetries in flavor-specific final states, which does not require
time dependent measurement. DCPV was recently observed
\cite{AKpiAKpi0} in $B^0\to K^+\pi^-$ decay, i.e. ${\cal A}_{K^+\pi^-}
= -0.115 \pm 0.018$. Having similar dominating penguin and tree
contributions, one would naively expect that ${\cal A}_{K^+\pi^-} =
{\cal A}_{K^+\pi^0}$. However, no indication of DCPV was seen in
charged $B^+\to K^+\pi^0$, i.e.  ${\cal A}_{K^+\pi^0} = 0.04 \pm
0.04$. The difference with ${\cal A}_{K^+\pi^-}$ could be due to an
enhanced color-suppressed amplitude $C$~\cite{LargeC}, or electroweak
penguin $P_{\rm EW}$ effects \cite{BurasEWP,Barger,Baek,Kpi0HNS}. The
former requires $C$ to effectively cancel the SM phase in
color-allowed tree amplitude $T$, without recourse to NP. For the
latter, NP CPV phases would be needed in $P_{\rm EW}$.

It would be intriguing if the two hints of NP, one in $\Delta{\cal
S}_f \neq 0$, the other in ${\cal A}_{K^+\pi^0} - {\cal
A}_{K^+\pi^-} \neq 0$, could be manifestations of the same NP
source. Since $\pi^0$ and $\phi$ (but not $\pi^-$) can materialize
from a virtual $Z$, the $B\to \pi^0K^0$, $\phi K^0$ modes are
sensitive to $Z$ penguins. The effect of a NP phase in $P_{\rm
EW}$ on $\Delta {\cal S}_f$, among several NP scenarios, was
studied in Ref.~\cite{BHNR}. In another work, some of us have
shown~\cite{Kpi0HNS} that the 4th generation could provide a
solution to the ${\cal A}_{K^+\pi^0} - {\cal A}_{K^+\pi^-} \neq 0$
problem through the electroweak penguin. The 4th generation can
make specific impact on $P_{\rm EW}$ because the $t'$ quark, like
the SM top, enjoys nondecoupling in $P_{\rm EW}$ \cite{HSW}, but
largely decouples from photonic and gluonic penguins. Furthermore,
it can provide a new CPV phase \cite{AH2003} through
$V^*_{t's}V_{t'b} \equiv r_{sb} e^{i\phi_{sb}}$.

In this work we show that the fourth generation can, for the right
choice of $\phi_{sb}$, give the correct trend for $\Delta {\cal
S}_f$ in $B^0\to \pi^0 K^0$ and $\phi K^0$, and is robust against
hadronic uncertainties. In contrast, we find $\Delta{\cal
S}_{\eta^\prime K^0}$ is largely diluted by hadronic effects that
are needed to account for the large rate.

For relevant 4th generation parameters, we take~\cite{Kpi0HNS}
\begin{equation}
m_{t'} = 300\ {\rm GeV}, \ \ r_{sb} \simeq 0.025,
\end{equation}
and vary $\phi_{sb}$ phase. Eq.~(2) is consistent with $b\to
s\ell^+\ell^-$ and $B_s$ mixing constraints~\cite{Kpi0HNS}. Larger
$m_{t'}$ or $r_{sb}$ \cite{rescale} could lead to larger effects
on $\Delta S_f$, but could run into trouble with the other $b\to
s$ constraints. To study (factorization) model dependence, we
compare results in naive factorization (NF)~\cite{NF}, QCD
factorization (QCDF)~\cite{QCDF,BN} and PQCD~\cite{PQCD}. We
further use QCDF to illustrate hadronic uncertainties. We choose
to use QCDF and PQCD circa 2003 because, in part stimulated by the
$\Delta{\cal S}$ and ${\cal A}_{K^+\pi^0}$ problems, these
factorization models are still being refined.


We adopt QCDF as our reference framework. Defining $\lambda_i
\equiv V^*_{is}V_{ib}$, one has $\lambda_u + \lambda_c + \lambda_t
+ \lambda_{t'} = 0$ with existence of $t^\prime$. To good
approximation, $\lambda_u$ is negligible compared with $\lambda_c
\simeq 0.04$, where we have taken the convention to keep $V_{cb}$
real, and placing the 3 CPV phases in $V_{ub}$, $V_{t's}$ and
$V_{t'd}$~\cite{HSS,HNS-KL}. This makes clear correspondence to
the standard phase convention for 3 generation case. The unitarity
condition $\lambda_t \simeq -\lambda_c - \lambda_{t'}$ allows one
to absorb the $t$ effect into the $\lambda_c$ dependent part (SM
term), and the NP $\lambda_{t'}$ dependent part that respects GIM
\cite{AH2003,Kpi0HNS}.

The $\overline B^0\to \pi^0\overline K^0$ amplitude is
\begin{eqnarray}
{\cal M}_{\pi^0\overline K^0} \propto f_\pi F_{BK} ( \lambda_u
a_2^u + \frac{3}{2}\lambda_c \alpha_{3,{\rm EW}}^p -
\frac{3}{2}\lambda_{t^\prime}\Delta\alpha_{3,{\rm EW}}^p) \
\nonumber \\
 - f_K F_{B\pi}[ \lambda_c ( \alpha_{4}^p
                     - \frac{1}{2}\alpha_{4,\rm{EW}}^p
                     + \beta_{3}^{p} ) +
                       \frac{1}{2}\lambda_{t^\prime}\Delta\alpha_{4,{\rm EW}}^p
               ], \ \;
\end{eqnarray}
where $\alpha_{i(, {\rm EW})}^{p}$ and $\beta_3^p$ are defined in
Ref.~\cite{QCDF} and evaluated for the $\pi^0\overline K^0$ final
state, and $\Delta\alpha_{i,{\rm EW}}^p$ is the effective ($t$
subtracted) $t'$ contribution. For $\overline B^0\to \phi\overline
K^0$, there is no tree term, and one has
\begin{eqnarray}
{\cal M}_{\phi\overline K^0} & \propto & \lambda_c[ \alpha_{3}^p
+\alpha_{4}^p+\beta_{3}^p -\frac{1}{2}(\alpha_{3,{\rm EW}}^p
+\alpha_{4,{\rm EW}}^p+\beta_{3,{\rm EW}}^p)]
\nonumber \\
& & + \frac{\lambda_{t^\prime}}{2} (\Delta\alpha_{3,{\rm EW}}^p
+\Delta\alpha_{4,{\rm EW}}^p+\Delta\beta_{3,{\rm EW}}^p),
\end{eqnarray}
where $\alpha_{i,({\rm EW})}^{p}$ and $\beta_3^p$ are evaluated
for the $\phi\overline K^0$ final state. We have dropped the
common $f_\phi F_{BK}$ factor compared to Eq. (4), and we show
only the more important terms. The numerics was done with full
details according to Ref.~\cite{BN}.
The formula for $\overline B^0\to \eta'\overline K^0$ can be
analogously written, but is more elaborate which we do not reproduce
here.
We stress that the same expressions apply to the amplitudes in NF
framework as well, with the various coefficients taken at LO instead
of NLO.

In this work we estimate and quantify the impact of hadronic
uncertainties in QCDF. Among the hadronic parameters that enter
the decay amplitudes, three stand out as having the largest impact
due to uncertainties~\cite{QCDF}: the divergent part of the hard
spectator scattering integral $X_H$, the divergent part of the
weak annihilation integral $X_A$, and the first inverse moment of
the $B$ meson distribution amplitude $\lambda_B$. The first two
are estimated to be complex numbers of order $\ln (m_B/\Lambda_h)$
with $\Lambda_h=500$ MeV, and can therefore be parameterized
by~\cite{xhxa}
\begin{equation}
  X_{H,A} = \left(1 + \rho_{H,A}\, e^{i \phi_{H,A}}\right) \ln
  \frac{m_b}{\Lambda_h}.
  \label{XAXH}
\end{equation}
Our estimate of the hadronic uncertainties is based on the
variation of these parameters over a wide range as indicated in
Ref.~\cite{QCDF}. For reference, we also take as baseline
a ``standard'' scenario, in which we fix $\rho_H=0$, $\rho_A = 1$,
$\phi_A = -45^\circ$ and $\lambda_B=350$ MeV. This scenario
corresponds to the ``S3'' scenario of Ref.~\cite{BN}, although
small numerical differences in input parameters may lead to a
slight difference in final results~\cite{communication}. 



For $\overline B^0\to \pi^0\overline K^0$ in PQCD factorization,
we adopt the LO result used in Ref. \cite{Kpi0HNS}
\begin{eqnarray}
{\cal M}_{\pi^0\overline K^0} \propto \lambda_u f_\pi F_{ek} &+&
\lambda_c ( - f_K F_e^{P} - f_B F_a^{P} + f_\pi F_{ek}^{P}) \nonumber \\
 &-& \lambda_{t^\prime}f_\pi \Delta F_{ek}^{P},
\end{eqnarray}
where $F_e^{P}$, $F_a^{P}$, $F_{ek}$ and $F_{ek}^{P}$ are the
strong penguin, strong penguin annihilation, color suppressed tree
and (color allowed) electroweak penguin contributions,
respectively. These factorizable contributions can be computed by
following Ref.~\cite{PQCD}, and are tabulated in
Ref.~\cite{Kpi0HNS}.

For $\overline B^0\to \phi\overline K^0$, we have
\begin{eqnarray}
{\cal M}_{\phi\overline K^0} \propto \lambda_c( f_\phi F_e^{P} +
f_B F_a^{P} ) - \lambda_{t^\prime}f_\phi \Delta F_{e}^{P},
\end{eqnarray}
where the $F_i^P$s are evaluated for $\phi \overline K^0$
\cite{PQCDphiK} and not the same as in Eq. (6). We have performed
only an approximate computation in this case. We assume that the
scale $t$, where the Wilson coefficients are evaluated, has a mild
dependence on the momentum fraction $x$ and the impact parameter
$b$ which is conjugate to the parton transverse momentum. The
amplitude $F_{e}^{P}$, which is obtained by integrating over the
variables $x$ and $b$, becomes then proportional to $a_e(t)$ with
\begin{eqnarray}
a_e(t) & = & C_3+\frac{C_4}{3}+C_4+\frac{C_3}{3}+
C_5+\frac{C_6}{3} \nonumber \\
& & -\frac{1}{2}\left(C_7+\frac{C_8}{3} +C_9+\frac{C_{10}}{3}
+C_{10}+\frac{C_9}{3}\right).
\end{eqnarray}
By knowing now the numerical value of $F_{e}^{P}$~\cite{PQCD} and
the Wilson coefficients in the SM calculated at $t=m_b$, one can
then calculate $\Delta F_{e}^{P}$ with
\begin{equation}
\Delta F_{e}^{P} = F_e^P\left(\frac{a_e^{NP}}{a_e^{SM}}-1\right).
\end{equation}
The same procedure is not possible for $F_{a}^{P}$ and we keep
only the SM contribution by assuming $\Delta
F_{a}^{P}=0$~\cite{deltafap}. For $\eta'K$ mode, not much work has
been done in PQCD.

\begin{figure}[t!]
\smallskip  
\includegraphics[width=1.5in,height=1.0in,angle=0]{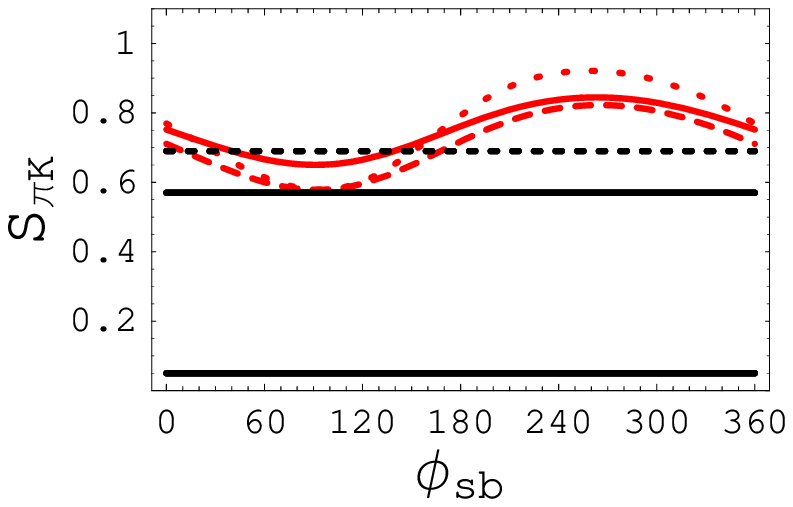}
\hskip0.2cm
\includegraphics[width=1.5in,height=1.0in,angle=0]{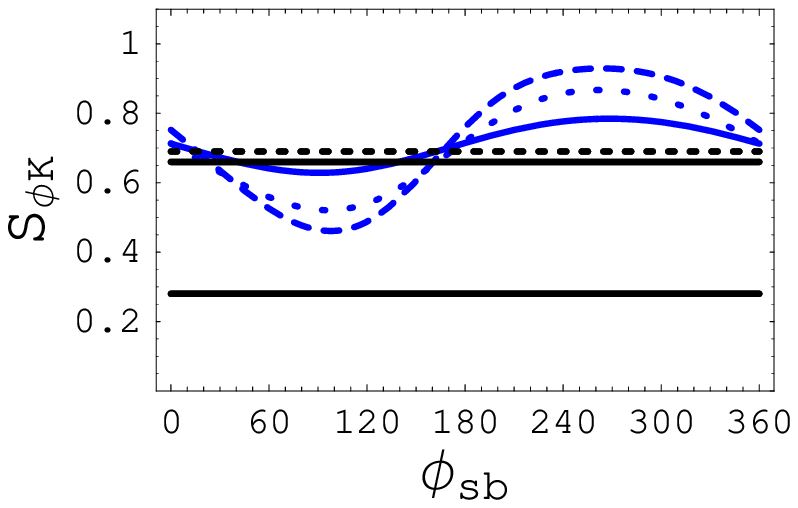}
\vskip1.1cm
 \caption{
  (a) $S_{\pi^0 K^0}$,
  (b) $S_{\phi K^0}$ vs $\phi_{sb}$ for QCDF at NLO with ``S3''
  parameters (solid), PQCD at LO (dashed), and NF (dots), which is
  QCDF at LO. The horizontal solid band is the current experimental
  range. 
 \label{fig:Fig1}
}
\end{figure}

To study the model dependence in different factorization approaches,
we plot $S_{\pi^0 K^0}$ and $S_{\phi K^0}$ vs $\phi_{sb}$ in Fig.~1
for QCDF at NLO (with ``S3''parameters), PQCD at LO, and NF. The
latter is far from realistic (for rates) and is just for
comparison. We see that in all three models, $S_{\pi^0 K^0}$ and
$S_{\phi K^0}$ dip below $\sin2\phi_1/\beta$ for $\sin\phi_{sb}
\gtrsim 0$, especially around $\phi_{sb} \sim
90^\circ$
Indeed, for a given size of NP contribution, a choice of a maximal
weak phase of 90$^\circ$ (or 270$^\circ$) tends to maximize the NP
effect on CPV while minimizing the NP effect on BR. It is interesting
to note that this is precisely what is needed for the 4th generation
to help resolve \cite{Kpi0HNS} the ${\cal A}_{K^+\pi^0} - {\cal
  A}_{K^+\pi^-} \neq 0$ problem.
Independently, $\phi_{sb} \sim 90^\circ$ is also the parameter
space where $b\to s\ell^+\ell^-$ and $\Delta m_{B_s}$ constraints
are best evaded \cite{Kpi0HNS,AH2003}. For $\phi_{sb} \sim
270^\circ$, although the $b\to s\ell^+\ell^-$ and $\Delta m_{B_s}$
constraints can still be tamed, both ${\cal S}_{\pi^0 K^0,\phi
K^0}$ and ${\cal A}_{K^+\pi^0}$ would be in disagreement with
experiment.

\begin{figure}[t!]
\includegraphics[width=1.5in,height=1.0in,angle=0]{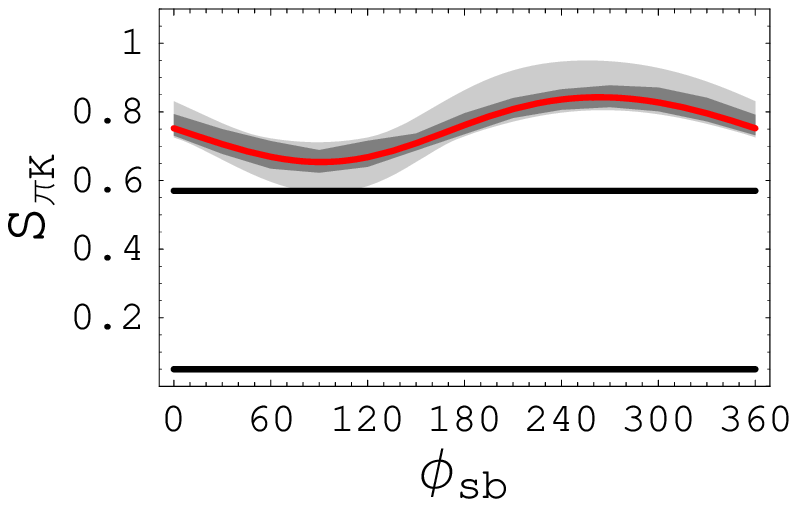}
\hskip0.2cm
\includegraphics[width=1.5in,height=1.0in,angle=0]{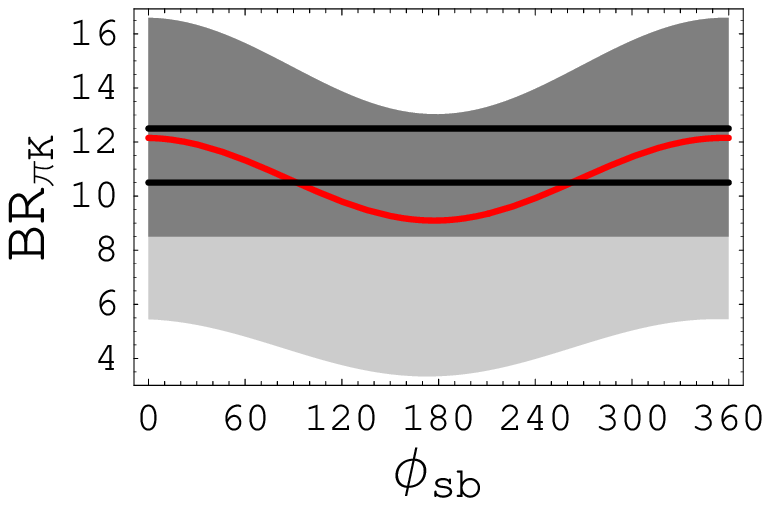}
\includegraphics[width=1.5in,height=1.0in,angle=0]{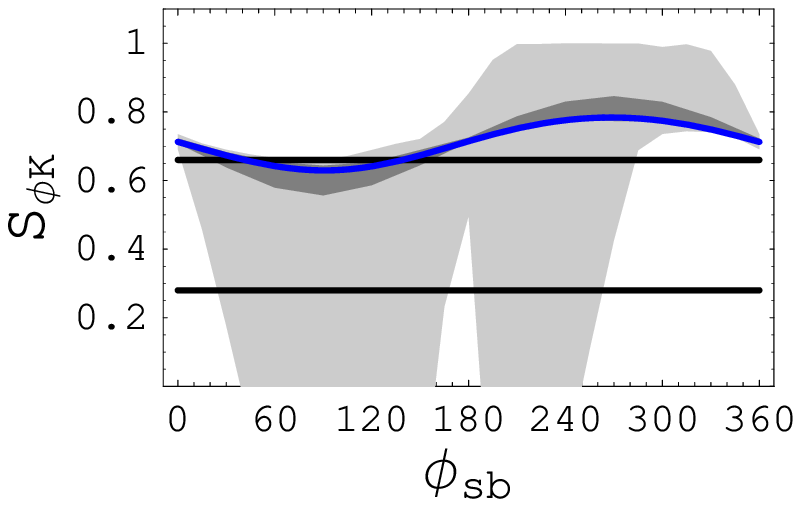}
\hskip0.2cm
\includegraphics[width=1.5in,height=1.0in,angle=0]{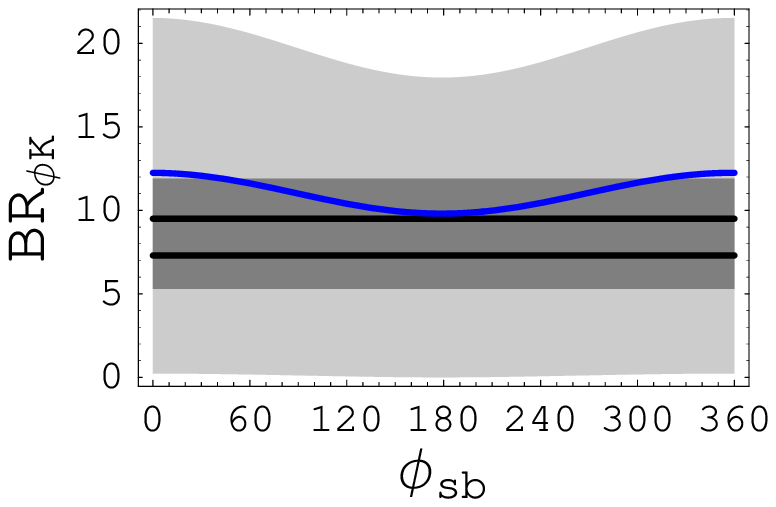}
\includegraphics[width=1.5in,height=1.0in,angle=0]{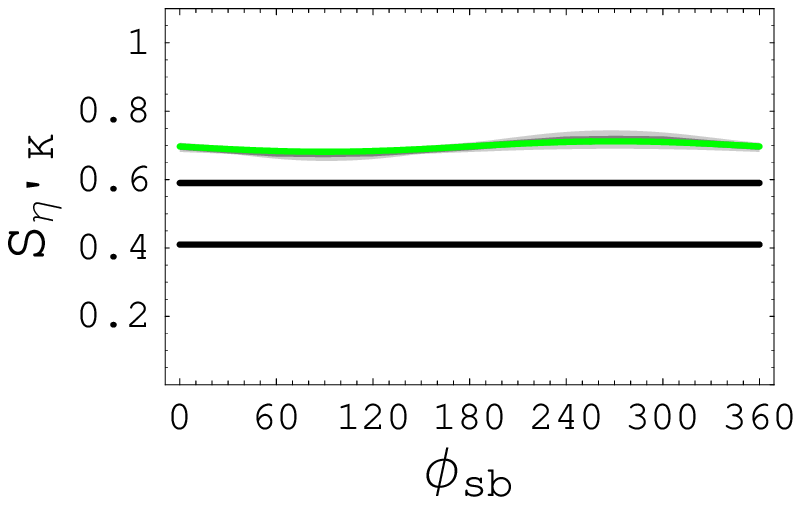}
\hskip0.2cm
\includegraphics[width=1.5in,height=1.0in,angle=0]{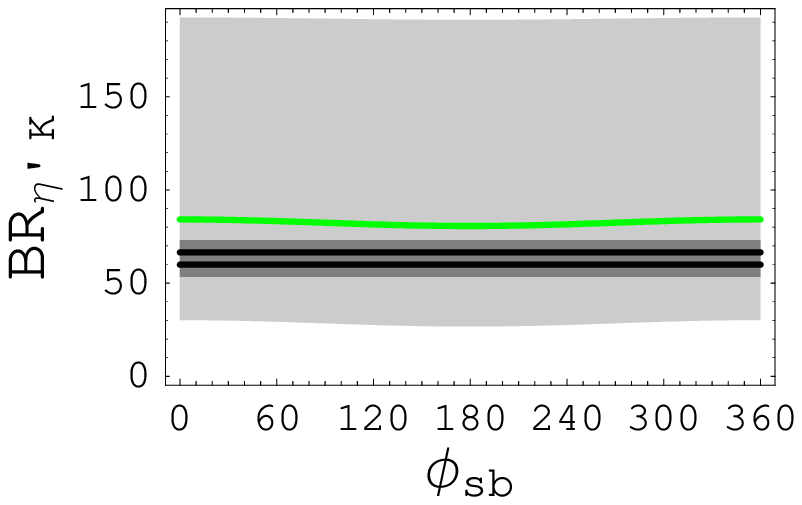}
\vskip0.5cm
 \caption{
 (a) $S_{\pi^0 K^0}$,        (b) ${\cal B}(\pi^0K^0)$,
 (c) $S_{\phi K^0}$,        (d) ${\cal B}(\phi K^0)$,
 (e) $S_{\eta^\prime K^0}$, (f) ${\cal B}(\eta^\prime K^0)$ vs $\phi_{sb}$,
 in QCDF at NLO. The curves are for the ``S3'' hadronic parameter
 settings. The light gray regions correspond to varying the hadronic
 parameters $X_A$, $X_H$ and $\lambda_B$ over the range indicated in
 Eq.~(10). The dark gray regions are obtained by varying the
 hadronic parameters over the same range as above, but keeping only
 the values for which the branching ratios are within $3\,\sigma$ of
 the experimental central values.  }
 \label{fig:Fig2}
\end{figure}

Hadronic parameters such as strong phases easily affect branching
ratios and DCPV asymmetries. Strong phases are definitely present in
$B\to K\pi$ decay as evidenced by the sizable ${\cal A}_{K^+\pi^-} =
-0.115 \pm 0.018$. The $S_f$ parameter, however, measures the weak
phase of the decay amplitude, and is less affected by hadronic
parameters~\cite{BHNR}. As mentioned, we illustrate this point, by
varying the hadronic parameters of QCDF at NLO around the ``S3''
scenario settings. In particular, we vary~\cite{QCDF,BN}
\begin{equation}
  \begin{array}{c}
    \rho_{A,H} \in (0,1), \ \phi_{A,H} \in [0,2\pi], \medskip\\
    \lambda_B \in [200,500]{\rm MeV},
  \end{array}
\end{equation}
for the $X_A$, $X_H$ and $\lambda_B$ parameters.

We plot $S_{\pi^0 K^0}$, $S_{\phi K^0}$ and $S_{\eta' K^0}$ vs
$\phi_{sb}$ in the left side of Fig.~2.
The light shaded regions correspond to varying the parameters over
the whole range indicated in Eq.~(10). The dark shaded regions
correspond to varying the hadronic parameters over the same range,
but keeping, for each mode, only the values that produce a
branching ratio (right side of Fig.~2) within $3\,\sigma$ of the
experimental central value. One sees that, indeed, the branching
ratios are strongly affected by the hadronic parameters, and most
of the hadronic parameter space cannot survive the bulk of rate
and DCPV data when considered together. In contrast, the range of
variation for $S_f$ is much more subdued.  This is encouraging:
the NP effect in $S_f$ for the $\pi^0 K^0$, $\phi K^0$ and $\eta'
K^0$ modes is robust.

We note that the effect of hadronic parameters, when varied over the
whole range, is rather strong for $S_{\phi K^0}$. However, when the
experimental constrains on the $B\to \phi K^0$ branching ratio are
taken into account, the hadronic uncertainty in $S_{\phi K^0}$ is
highly diminished.

Note, also, that $S_{\eta' K^0}$ gets strongly diluted away.
The reason behind this is the rather large rate of $B\to \eta'K$
decay, which seemingly draws from $CP$ conserving (``hadronic")
effects, since there is little evidence for CPV i.e. ${\cal
A}_{\eta'K} \simeq 0$ \cite{HFAG}. Furthermore, the $Z$ penguin
contribution has relatively small strength.
We believe the dilution of $S_{\eta' K^0}$ is a generic effect,
that is, it is very hard for NP CPV effects to shine through the
large hadronic effects, and $S_{\eta' K^0} \simeq
\sin2\phi_1/\beta$ should be expected. In this respect, the Belle
result of ${\cal S}_{\eta^\prime K^0} = 0.62\pm 0.12 \pm 0.04$,
which is fully consistent with $\sin2\phi_1/\beta = 0.69 \pm
0.03$, is easier to explain in most NP models \cite{etapK}. If the
BaBar result holds out eventually, it would need some conspiracy
between NP and hadronic effects to realize theoretically.

\begin{figure}[t!]
\includegraphics[width=1.5in,height=1.0in,angle=0]{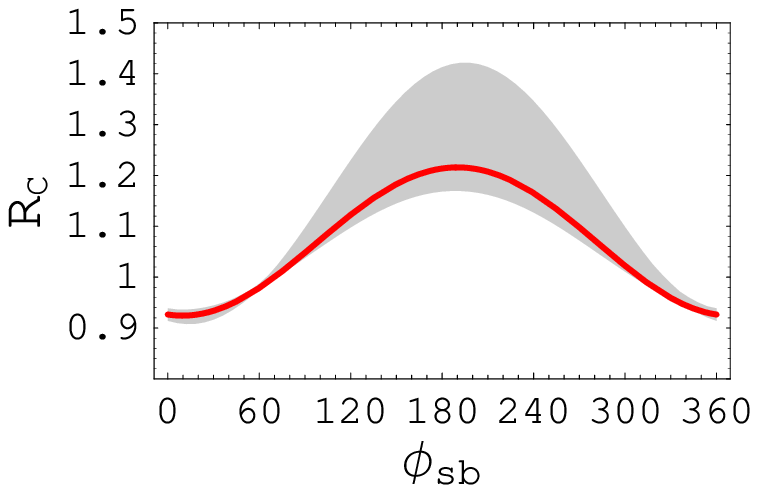}
\hskip0.2cm
\includegraphics[width=1.5in,height=1.0in,angle=0]{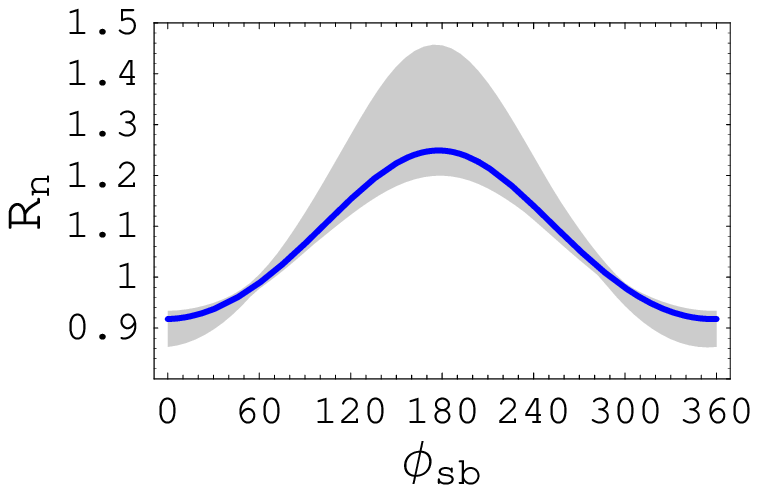}
\vskip0.2cm
\caption{ Ratio of branching ratios (a) $R_c$ and (b) $R_n$ for $B^+$
  and $B^0$ decay to $K\pi$, as defined in Ref.~\cite{BurasEWP}, vs
  $\phi_{sb}$, for QCDF at NLO. The curves and the light gray regions
  are obtained as in Fig.~2.  }
 \label{fig:Fig3}
\end{figure}

We offer some remarks before closing.
We have studied the ratio of branching ratios $R_c$, $R_n$ and $R$,
which are for $B^+$, $B^0$, and the lifetime corrected $K^+\pi^-$ over
$K^0\pi^+$ ratio, respectively. Indeed, these rate ratios are
attractive in that they suffer considerably less hadronic
uncertainties. We plot $R_c$ and $R_n$ vs $\phi_{sb}$ in Fig.~3 for
QCDF at NLO and varying hadronic parameters over the full range of
Eq.~(10).
The contrast with the branching ratio plot in Fig.~2 is striking.
Interestingly, for $|\phi_{sb}| \lesssim 80^\circ$, the hadronic
uncertainties are even less significant (a bit more for $R$), and
the results with 4th generation are basically consistent with
experiment. But for $\phi_{sb} \sim \pi$, besides much larger
hadronic uncertainties, $R_c$ and $R_n$ would deviate
substantially (being larger) from data, and disallowed. This is
again consistent with the analysis from ${\cal A}_{K^+\pi^0} -
{\cal A}_{K^+\pi^-}$ as well as $b\to s\ell^+\ell^-$ and $B_s$
mixing, and with our findings for $S_{\pi K}$ and $S_{\phi K}$.

$S_f$ has been studied experimentally in quite a few other modes
such as $f = f_0(980)K_S$, $\omega K^0$~\cite{HFAG}, as well as
3-body modes such as $f = K\bar KK$ and $K_S\pi^0\pi^0$. The
interest in $S_{\eta K}$, $S_{\omega K}$ and $S_{\rho^0 K}$ have
been stressed \cite{Beneke, BHNR}. We have studied these modes and
found the effect of hadronic uncertainties to be more significant.
Thus, experimental studies in these modes would shed little light
on NP parameters, except that $S_{\rho^0 K} > \sin2\phi_1/\beta$
is likely realized. The theory for 3-body modes is even less
developed.
Similarly, DCPV depends sensitively on hadronic phases, and much
theoretical work is currently ongoing to elucidate these. We
therefore leave this for future studies. Our studies do show that
DCPV in the above mentioned 2-body modes are in general consistent
with data, since experimental errors are still large. The only
firmly measured DCPV is in ${\cal A}_{K^+\pi^-}$, while
Ref.~\cite{Kpi0HNS} has demonstrated that the 4th generation may
help resolve the ${\cal A}_{K^+\pi^0} - {\cal A}_{K^+\pi^-} \neq
0$ problem.

Finally, we note from Fig.~2 that for QCDF the experimental central
values are unattainable once the branching ratio is constrained to
within 3$\sigma$ of experiment. (Note, however, from Fig.~1, that our
approximate PQCD result could fit the $S_{\phi K}$ central value.)  If
the experimental central values for $S_{\phi K}$, $S_{\pi K}$ and
$S_{\eta' K}$ persist, more work on factorization models seem needed
to shed further light on whether the 4th generation, or other New
Physics, could account for the observed effect.

In summary, we have studied in this work the effect of a 4th
generation model on the TCPV parameter $S_f$ for $f=\pi K^0$, $\phi
K^0$ and $\eta' K^0$. We have shown, using QCDF at NLO, that the NP
effects on these $S_f$'s are rather robust against hadronic
uncertainties. This robustness may be generic to a large class of NP
models. We found that the same 4th generation parameters that explain
${\cal A}_{K^+\pi^0} \sim 0$ while ${\cal A}_{K^+\pi^-} \simeq -11\%$,
can give the correct trend in $S_f$.  However, we also showed that
$S_{\eta' K^0}$, $S_{\pi K^0}$ and to a lesser degree $S_{\phi K^0}$
are predicted to be closer to $\sin2\phi_1/\beta$ than the current
data indicate. Due to the robustness of the $S_f$, better measurements
could provide an important test of the 4th generation model as well as
other NP models.

\vskip 0.3cm \noindent{\bf Acknowledgement}.\ This work is
supported in part by NSC-94-2112-M-002-035 and
NSC94-2811-M-002-053 of Taiwan, and HPRN-CT-2002-00292 of Israel.
We would like to thank M. Beneke and Y. Nir for very
useful discussions.

%


\begin{thebibliography}{99}
\bibitem{etapK} The BaBar value of ${\cal S}_{\eta^\prime K^0} = 0.36
  \pm 0.13 \pm 0.03$, and the Belle value of ${\cal S}_{\eta^\prime
    K^0} = 0.62\pm 0.12 \pm 0.04$ are in some variance. One could
  inflate the error of the mean by a factor of $\sqrt{2}$.  We refrain
  from doing so and use the value quoted by HFAG, for sake of
  consistency with other modes. Note that the Belle value is in good
  agreement with $\sin2\phi_1$.
\bibitem{rescale}
Choosing a larger $m_{t'}$ can be compensated by a smaller
$r_{sb}$, and vice versa, for the same effect.
\bibitem{xhxa}
In principle both $X_H$ and $X_A$ are mode
  dependent. However, since we vary them over the whole possible range
  in any case, this point does not affect our analysis.
\bibitem{deltafap}
Since one expects very strong GIM cancellation between $t'$ and
$t$ for strong penguin, the approximation should be a very good
one.
\end{thebibliography}
\end{document}